\documentclass[11pt]{article}
\usepackage{geometry}  
\usepackage[T1]{fontenc}  % Standard package for selecting font encodings
\usepackage[utf8x]{inputenc}  % Accept different input encodings
\usepackage{lineno}  % Line numbers on paragraphs
\usepackage{fancyhdr}  % Extensive control of page headers and footers in LaTeX2ε
\usepackage{enumitem}  % Control layout of itemize, enumerate, description
\usepackage{hyperref}  % Extensive support for hypertext in LaTeX
\usepackage{titling}  % Control over the typesetting of the \maketitle command
\usepackage[numbers]{natbib}  % Flexible bibliography support
\usepackage{mathtools}  % Mathematical tools to use with amsmath
\usepackage{titlesec}  % Select alternative section titles
\usepackage{lastpage}  % Reference last page for Page N of M type footers
\usepackage[english]{babel}  % Multilingual support for Plain TeX or LaTeX
\usepackage{amsmath}  % AMS mathematical facilities for LaTeX
\usepackage{amsfonts}  % TeX fonts from the American Mathematical Society
\usepackage{amssymb}  % Additional symbols from American Mathematical Society
\usepackage{wasysym}  % LaTeX support file to use the WASY2 fonts
\usepackage{bbm}  % "Blackboard-style" cm fonts
\usepackage{array}  % Extending the array and tabular environments
\usepackage{xr}  % References to other LaTeX documents
\usepackage{float} % Improved interface for floating objects
\usepackage{color}% For colored text
\usepackage{algorithm, caption}
\usepackage{algorithmic} % Algorithm support package
\usepackage{xspace} % helps to space algorithmic environment 
\usepackage{booktabs}
\usepackage{multirow}
\usepackage{stmaryrd}
\usepackage{mathrsfs}
\usepackage{siunitx} % align in decimal point
\usepackage{textcomp} % trademark symbol
\usepackage{hyperref} % url
\usepackage{changes}

\usepackage{subcaption}
\usepackage{graphicx}
\usepackage[normalem]{ulem}
\usepackage{soul}
\hypersetup{%
  pdfborderstyle={/S/U/W 1}% border style will be underline of width 1pt
}

\setcitestyle{round,numbers}
\makeatletter
\def\BState{\State\hskip-\ALG@thistlm}
\makeatother

\titleformat{name=\section}{\normalfont\LARGE\bfseries}{}{0pt}{}
\titleformat{name=\subsection}{\normalfont\Large\bfseries}{}{0pt}{}
\titleformat{name=\subsubsection}{\normalfont\large\bfseries}{}{0pt}{}

\newcommand{\email}[1]{\href{mailto:{#1}}{{#1}}}

\newcommand{\keywords}[1]{\textbf{Keywords}: {#1}}

\newcommand{\wordcount}[2]
{\begin{tabular}{rl}%
\textbf{Manuscript word count}:     & \textr{}{4,086}\\
\textbf{Abstract word count}:         & 
{190}\\
\end{tabular}}

\newcommand{\optincludegraphics}[2][]{}
\newcommand{\optinput}[1]{}
\newtagform{brackets}{[}{]}
\usetagform{brackets}

\newcommand{\thejournal}[1]{Magnetic Resonance in Medicine}

 \newcommand{\textr}[1]{\textcolor{red}{#1}}

 \newcommand{\Real}{{\mathbb{R}}}

\title{Automatic image-based respiratory signal extraction in real-time CMR}
\immediate\write18{texcount -sum=1,1,1,1,1,1,1 -merge -incbib -dir -utf8 \jobname.tex > \jobname.wcTotal }
\immediate\write18{texcount -sub=none -merge -incbib -dir -utf8 \jobname.tex | grep _main_ | grep -oE '[0-9]+' | python -c "import sys; print(sum(int(l) * w for l, w in zip(sys.stdin, [1, 1, 0, 0, 0, 0, 0])))" > \jobname.wcManuscript }
\immediate\write18{texcount -sub=none -merge -incbib -dir -utf8 \jobname.tex | grep Abstract | grep -oE '[0-9]+' | python -c "import sys; print(sum(int(l) * w for l, w in zip(sys.stdin, [1, 1, 0, 0, 0, 0, 0])))" > \jobname.wcAbstract }

\newcommand{\wcTotal}{\clearpage{\noindent\large{\bf Detailed Word Count} (not to be included for submission)}\verbatiminput{\jobname.wcTotal}}
\newcommand{\wcManuscript}{\input{\jobname.wcManuscript}}
\newcommand{\wcAbstract}{\input{\jobname.wcAbstract}}
\begin{document}

% ======================================================================
%TC:ignore
\begin{titlepage}
\begin{center}
\sf{Submitted to Magnetic Resonance in Medicine}\\
\bigskip
{\noindent\LARGE\bf \thetitle}
\end{center}

\bigskip

% : Insert author names, affiliations and corresponding author email
% : (do not include titles, positions, or degrees).
%FIXME
\begin{center}\large
    Chong Chen,\textsuperscript{1}
    Yingmin Liu,\textsuperscript{2} 
	Orlando P. Simonetti,\textsuperscript{2,4,5}
	Rizwan Ahmad\textsuperscript{1,2,3}
\end{center}

\bigskip

\noindent
%FIXME
\begin{enumerate}[label=\textbf{\arabic*}]
\item Biomedical Engineering, The Ohio State University, Columbus OH, USA
\item Davis Heart \& Lung Research Institute, The Ohio State University, Columbus OH, USA
\item Electrical and Computer Engineering, The Ohio State University, Columbus OH, USA
\item Internal Medicine, The Ohio State University, Columbus OH, USA
\item Radiology, The Ohio State University, Columbus OH, USA
\end{enumerate}

\bigskip

% : Use the dagger symbol to denote a single equal contribution authorship.
% : Multiple equal-contribution authorship may be included in the acknowledgments.
%FIXME
%\textbf{{†}}: These authors contributed equally to this work.

% : Use the asterisk to denote corresponding authorship.
% : Provide email address in note below.
%FIXME
\textbf{*} Corresponding author:

\indent\indent
\begin{tabular}{>{\bfseries}rl}
Name        & Chong Chen                                                    \\
Department    & Biomedical Engineering                                        \\
Institute    & The Ohio State University                                        \\
Address     & 460 W 12th Ave, Room 311                                        \\
            & Columbus OH 43210, USA                                        \\
E-mail        & \email{chen.7211@osu.edu}                                        \\
\end{tabular}

\vfill

% ======================================================================
% : set word count results (+++ must be included, --- must be excluded)
%     +++ introduction, theory, methods, results, discussion, conclusion,
%        appendix, 
%     --- title page, abstract, figure captions, tables, table captions,
%        references, revision markings
% : first argument is the manuscript word count
% : second argument is the abstract word count
% : to use `texcount` results, use '%TC:ignore'/'%TC:endignore' directives.
% : \wcManuscript and \wcAbstract should perform the correct word count.
% : display detailed word count
\wordcount{\wcManuscript}{\wcAbstract}

%\indent\indent
%\begin{tabular}{>{\bfseries}rl}
%    Manuscript Word Count:    & \hl{TBD}\\
%    Abstract  Word Count:    & \hl{TBD} \\
%\end{tabular}

\end{titlepage}
%TC:endignore
% ======================================================================

% ======================================================================
% ======================================================================
\pagebreak
% ======================================================================
% ======================================================================

% ======================================================================
%TC:break Abstract
\begin{abstract}
%FIXME

\noindent
\textbf{Purpose:} 
To develop a fully automatic method for extraction and directionality determination of respiratory signal in free-breathing, real-time (RT) cardiac MRI.
\\
\textbf{Methods:}
The respiratory signal is extracted by a principal component analysis method from RT cine images. Then, a two-step procedure is used to determine the directionality (sign) of the respiratory signal. First, the signal polarity of all slices is made consistent with a reference slice. Second, a global sign correction is performed by maximizing the correlation of the respiratory signal with the zeroth-moment center curve. The proposed method is evaluated in multi-slice RT cine from eleven volunteers and two patients. The motion in a manually selected region-of-interest is used as reference.
\\
\textbf{Results:} 
 The extracted respiratory signal using the proposed method exhibits high, positive correlation with the reference in all cases and is more robust compared to a recently proposed method. To demonstrate its clinical utility, the method is used to identify heartbeats from inspiratory and expiratory phases and, in turn, to account for respiration-induced changes in the cardiac output.%Also, it can accurately identify heartbeats from expiratory and inspiratory phases.
 %The clinical impact of this work is demonstrated by showing that the proposed method can identify heartbeats from expiratory and inspiratory phases that, in turn, impact cardiac output.% varies with the respiratory phase, which can be automatically determined using the proposed method.
\\
\textbf{Conclusions:} 
The proposed method enables fully automatic extraction and directionality determinations of respiratory signal from RT cardiac cine images, allowing accurate cardiac function quantification.

\end{abstract}
% ======================================================================
% : set search-engine keywords (3 to 6)
\bigskip

%TC:break _main_
% ======================================================================
% ======================================================================
\pagebreak
% ======================================================================
% ======================================================================

\section{INTRODUCTION}
Cardiovascular magnetic resonance imaging (CMR) is the gold standard to evaluate cardiac morphology and ventricular function. To reduce the influence of respiratory and cardiac motion, breath-held (BH) segmented k-space acquisitions are widely used in CMR. However, such protocols can be challenging for patients with poor respiratory control or arrhythmia. Moreover, they are unable to capture the influence of respiratory motion on the cardiac output. As a result, there is a growing demand to develop real-time (RT) CMR methods that are insensitive to arrhythmia and respiratory motion. 
%\textr{Respiratory motion is one of the main sources of artifacts and inaccurate quantification in cardiac imaging. To reduce its influence, breath-held (BH) segmented k-space acquisitions are widely used. However, such protocols have limitations for patients with arrhythmia or poor respiratory control. Moreover, they are unable to capture respiratory motion-induced changes in the cardiac output. As a result, there is a growing demand to develop real-time (RT) cardiovascular magnetic resonance imaging (CMR) methods that are insensitive to arrhythmia and respiratory motion.}
%
%Cardiac cine, i.e., creating a movie of the beating heart, is one of the most common applications of CMR.}
%Respiratory motion is one of the main sources of artifacts and inaccurate quantification in cardiac imaging. To reduce its influence, breath-held (BH) segmented k-space acquisitions are widely used. However, such protocols have limitations for patients with arrhythmia or poor respiratory control. Moreover, they are unable to capture respiratory motion-induced changes in the cardiac output. As a result, there is a growing demand to develop real-time (RT) cardiovascular magnetic resonance imaging (CMR) methods that are insensitive to arrhythmia and respiratory motion.
%Cardiac cine, i.e., creating a movie of the beating heart, is one of the most common applications of CMR. 
In RT cardiac cine, 2D slices are collected sequentially under free-breathing conditions. Recently proposed advanced data acquisition and image reconstruction methods have allowed the spatial and temporal resolution of RT cine to rival that of the traditional BH segmented acquisition~\cite{feng2013highly,kellman2009high, chen2019sparsity}. After image reconstruction, one complete heartbeat from each slice can be used for cardiac function quantification. To get accurate and reproducible cardiac function measurements, all selected heartbeats must belong to a single respiratory phase, e.g., peak-expiration \cite{wu2015efficient, wu2017improved}. Right ventricular quantification, in particular, has been shown to be sensitive to the respiratory phase~\cite{claessen2014interaction}. 

Navigator echoes are commonly used for prospective compensation of respiratory motion~\cite{wang1996navigator}. Such techniques, however, do not allow imaging across the entire cardiac cycle and are not compatible with steady-steady free-precession sequences, which are commonly used for cardiac cine. External devices, such as belts or pillow bellows with pressure sensors, are also able to measure the respiratory signal~\cite{santelli2011respiratory}. However, these devices require prior setup, must be synchronized with the MRI acquisition, and do not work reliably in patients. Consequently, they are not commonly used in clinical practice. Self-gating has also been proposed for RT imaging, where the same k-space location is sampled repeatedly during the scan~\cite{larson2004self}. The resulting signal is then processed to extract cardiac and respiratory signals. For Cartesian sampling, however, this strategy requires modifying the pulse sequence and can lower the overall efficiency to accommodate the increased repetition time or the additional self-gating readouts~\cite{crowe2004automated}. Alternatively, the respiratory signal can be extracted directly from reconstructed RT images collected without scarifying acquisition efficiency. In the MRI literature, several image-based strategies for respiratory signal extraction have been reported. %~\cite{kellman2008fully, usman2015manifold}. 
For example, methods based on tracking lung air density, diaphragm motion, image center of mass, and image registration have been proposed~\cite{CTDiaphragm,MRIDiaphram,PETCOM,MRInonrigid}. However, in these methods, the image features with pronounced respiratory motion must be selected manually. Also, such features may not be consistently visible in all slices. %assumption that such and it is problematic if the organ related features do not show up in the image.
%Another method is tracking the , however, its performance is not stable, especially when there are opposite motions in different regions of the image or significant through-plane motion. 

Instead of tracking image features, several recently proposed methods rely on principal component analysis (PCA) for respiratory motion extraction. PCA-based methods map the high-dimensional image series to a low dimensional space, while preserving salient spatiotemporal information. Such methods have been successfully used to extract the respiratory signal from PET, CT, and MRI images~\cite{PETCTPCA,xRay_PCA,MRIPCA}. Extensions of PCA, such as kernel-PCA and Laplacian eigenmaps, have also been employed for respiratory signal extraction~\cite{X_Ray_kernal_PCA,SPECTmanifold,PETCTmanifold,usman2015manifold}. A major challenge for such methods is that the extracted signal has an uncertain directionality (i.e., arbitrary sign) with respect to the physical motion, which means that the maximal signal could refer to either inspiration or expiration. An inconsistent directionality assignment to respiratory signals from different slices in the cine stack can lead to a mixing of heartbeats from inspiration and expiration phases, which can impact cardiac function quantification. %inconsistent respiratory phases of the selected heat beats, and further inaccurate cardiac function quantification.
%Among them, PCA-based methods~\cite{thielemans2011device,novillo2019unsupervised} are particularly appealing because they are fast and independent of the slice orientation. However, due to the nature of PCA, the extracted signal has an arbitrary sign, which means that the % direction of the extracted motion is undefined and the 
%maximal signal could refer either to the inspiration or expiration phase. 
A simple approach to correctly assign directionality is to assume that more time is spent in expiration than inspiration~\cite{SPECTmanifold}. However, this assumption does not hold consistently across all subjects and is only applicable when there are several respiratory cycles available. To resolve this issue, more robust methods have been proposed. For example, Bertolli et al.~\cite{PET_sign_correction} resolved the directionality ambiguity in PET images by assuming the directionality of the respiration-induced cranio-caudal motion of the internal organs. For MRI, Novillo et al.~\cite{MRIPCA} recently proposed an unsupervised directionality determination method based on tracking the diaphragmatic motion or estimating the lung volume, but this method requires selecting the lung-diaphragm interface on cine image series.
%limiting its performance when the interface doesn't show up in the images.~\cite{novillo2019unsupervised}.

This study is an extension of our preliminary work \cite{chen2020automatic}, where a fully automatic method to extract directionality-resolved respiratory signal from a stack of RT cardiac cine was proposed and evaluated in healthy volunteers. In this work, we further optimize the algorithm, include additional data from patients, and make a direct comparison to the recently proposed method by Novillo et al.~\cite{MRIPCA}, which is referred to as Unsupervised Respiratory signal Extraction (URE). Also, to demonstrate clinical utility, we use the proposed method to segregate peak-expiration and peak-inspiration heartbeats from each slice and show the respiratory motion-induced variations in cardiac output in volunteers.

% This study is an extension of our preliminary work \cite{chen2020automatic}, where a fully automatic method to extract directionality-resolved respiratory signal from a stack of RT cardiac cine was proposed and evaluated in healthy volunteers. In this work, we further optimize the method, include additional data from patients, make comparison with the recently proposed method by Novillo et al.~\cite{MRIPCA}, which is referred to as Unsupervised Respiratory signal Extraction (URE).
%Linear PCA is applied to the preprocessed temporal image series to extract the respiratory motion. Then, we use a two-step procedure to determine the sign of the respiratory signal. First, a reference slice is selected and the signal polarity of the other slices is adjusted to be consistent with the reference. Then the overall sign correction is performed based on their correlation to the Location of the intensity Mean (LoM) curve. 
%Using data from eleven volunteers and two patients, we validate our method and compare it to the method . 
% To demonstrate clinical benefit, we use the proposed method to select one peak-expiration and one peak-inspiration heartbeat from each slice and then show the respiratory motion-induced variations in cardiac output in volunteers.

\section{THEORY}
%\textr{Move all the figures and tables towards the end, with one figure on one page. You may use 'clearpage' command at the end of each figure.}

\subsection{Extraction of the respiratory signal}
\begin{figure*}[htb]
  \centering
  \centerline{\includegraphics[width=16cm]{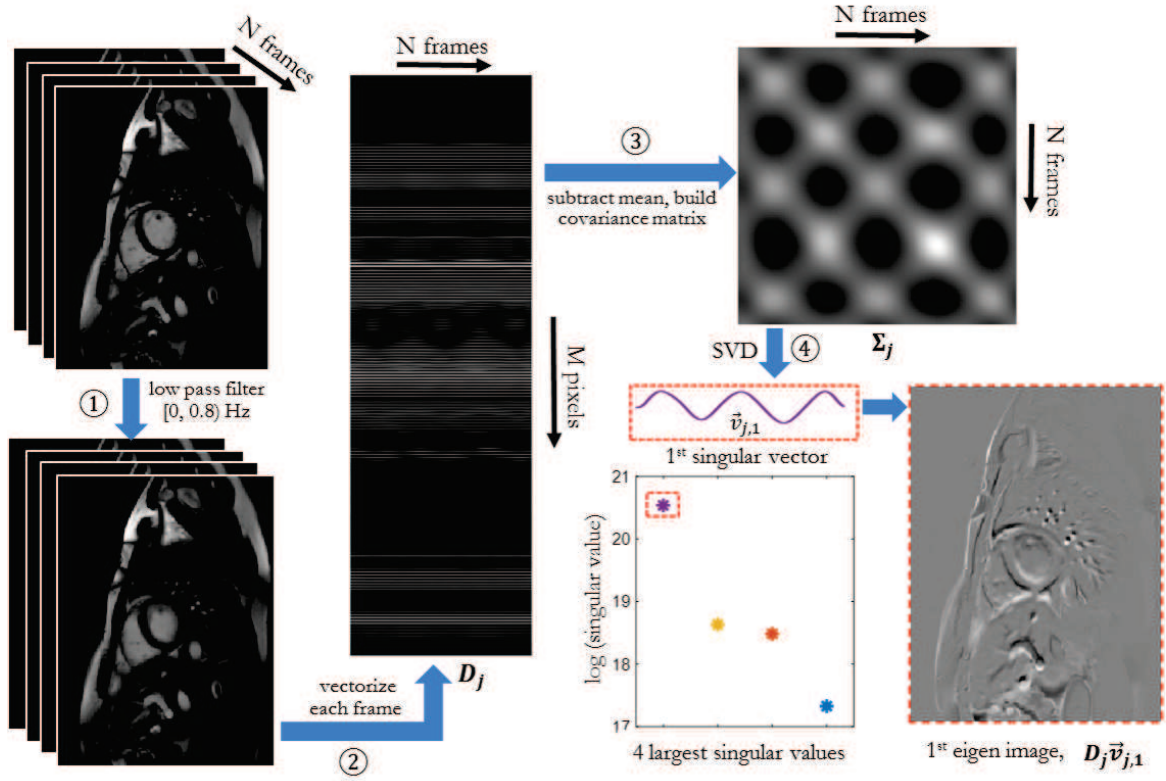}}
%  \vspace{2.0cm}
\caption{Extraction of the respiratory signal from the $j^\text{th}$ slice. First, a low-pass filter $[0-0.8\,\text{Hz}]$ is applied along the temporal dimension to suppress the cardiac motion. Second, the image series is flattened into a matrix, $\boldsymbol{D}_j$. Third, temporal mean is subtracted from $\boldsymbol{D}_j$ (Eq.~\ref{eq:D}), and the resulting matrix is used to compute the covariance matrix, $\boldsymbol{\Sigma}_j$ (Eq.~\ref{eq:cov}). Fourth, eigen decomposition is performed, with the first singular vector, $\Vec{v}_{j,1}$, corresponding to the respiratory signal. Finally, the $1^\text{st}$ ``eigen image'' is obtained by $\boldsymbol{D}_j\vec{v}_{j,1}$, which can be interpreted as a projection of the image series on the corresponding respiratory signal and is used in the directionality determination procedure.
}
\label{fig:method-extract-signal}
\end{figure*}

\noindent 
An overview of the extraction of the respiratory signal is given in Fig.~\ref{fig:method-extract-signal}. Consider an image series with $N$ frames and $M$ pixels per frame. First, we filter the images using a low pass filter $[0-0.8\,\text{Hz}]$ along the temporal dimension to suppress the higher frequency components, e.g., from the cardiac motion. After vectorizing each frame, 
the image series from the $j^\text{th}$ slice can be represented by matrix, $\boldsymbol{D}_j\in\mathbb{R}^{M\times N}$, with each column corresponding to a single frame. Then, we subtract the temporal mean from $\boldsymbol{D}_j$ and construct the covariance matrix $\boldsymbol{\Sigma}_j \in\mathbb{R}^{N\times N}$ using Eqs.~\ref{eq:D} and \ref{eq:cov}, i.e.,
   \begin{align}
   \label{eq:D}
   \boldsymbol{\tilde{D}}_j (m,n) &= \boldsymbol{D}_j (m,n) - \frac{1}{N} \sum_{n'=1}^{N} \boldsymbol{D}_j (m,n') \\
   \label{eq:cov}
\boldsymbol{\Sigma}_j &= \boldsymbol{\tilde{D}}_j^T \boldsymbol{\tilde{D}}_j
   \end{align}
\noindent 
The eigen decomposition of $\boldsymbol{\Sigma}_j$ is performed to obtain right singular vectors, $\{ \vec{v}_{j,1}, \vec{v}_{j,2},...,\vec{v}_{j,N} \}$, of $\boldsymbol{\tilde{D}}_j$. These singular vectors capture different motions present in $\boldsymbol{\tilde{D}}_j$, including the respiratory motion. Since the cardiac motion has been suppressed via low-pass filtering, the first singular vector, $\Vec{v}_{j,1}$, is expected to represent the respiratory signal. Lastly, the $1^\text{st}$ ``eigen image'' from the $j^\text{th}$ slice is obtained by $\boldsymbol{D}_j\vec{v}_{j,1}$, which can be interpreted as a projection of the image series on the corresponding respiratory signal, $\vec{v}_{j,1}$, and is used in the directionality determination procedure described below. For brevity, we drop the second subscript and represent the extracted respiratory motion in the $j^\text{th}$ slice by $\vec{v}_j$.
%Due to the nature of DR based method, the sign of the extracted respiratory signal $\Vec{v}_{j,1}$ is arbitrary. Sign correction procedure is needed to determine the end expiration (PE) and end inspiration (PI) respiratory phases.

\subsection{Directionality determination}
\begin{figure*}[htb]
  \centering
  \centerline{\includegraphics[width=18cm]{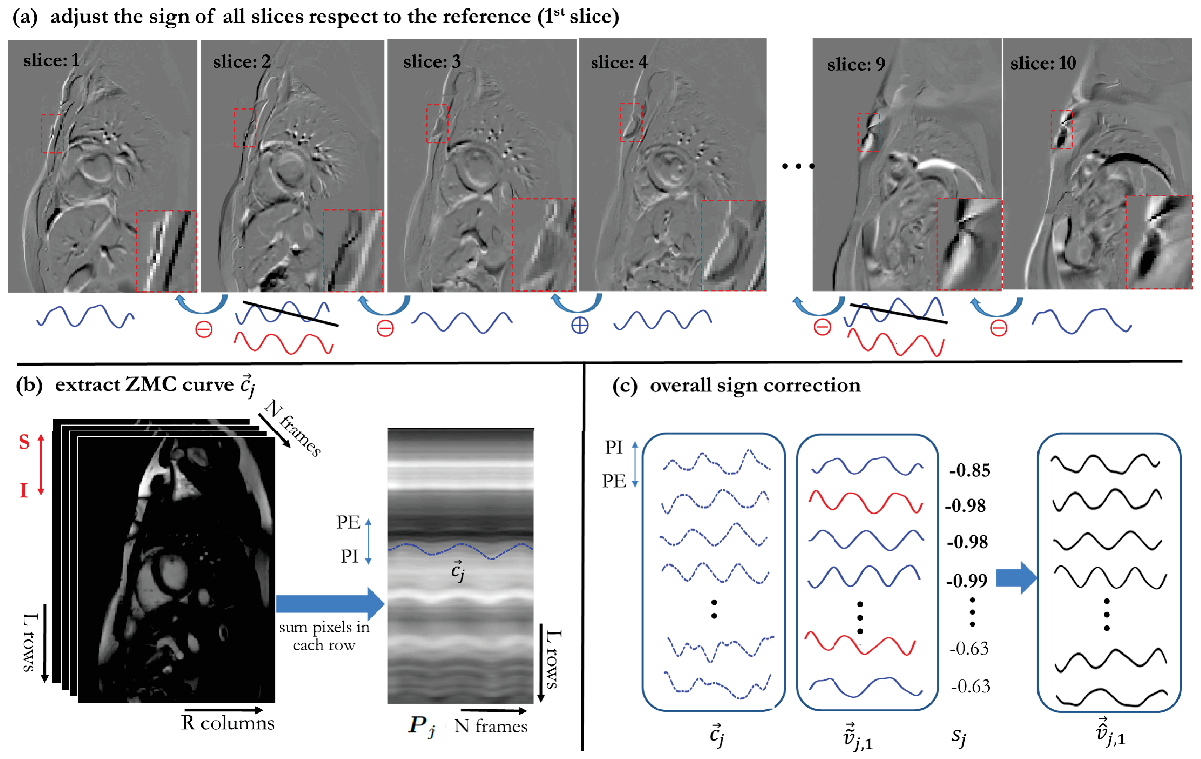}}
%  \vspace{2.0cm}
\caption{
Proposed directionality (sign) determination procedure. (a) Step 1: A pair-wise inter-slice correlations are used to fix the sign of individual slices with respect to the reference ($1^{\text{st}}$ slice). (b,c) Step 2: An overall sign correction is performed using the correlations, $s_j$, of slices with $\vec{c}_j$. Here, S and I represent superior and inferior directions, respectively, and PE and PI represent peak-expiration and peak-inspiration, respectively.}
\label{fig:method-sign-correction}
\end{figure*}

\noindent 
As Illustrated in Fig.~\ref{fig:method-sign-correction}, we adopt a two-step procedure to resolve directionality (sign) ambiguity in the respiratory motion from multi-slice RT cardiac cine. In the first step, we choose the $1^{\text{st}}$ slice ($j=1$) as reference and adjust the signs of all the other slices ($j=2,3,\dots,J$) with respect to the reference. To this end, we estimate Pearson correlation coefficient, $r_{j,j+1}$, between $\boldsymbol{D}_j\vec{v}_{j}$ and $\boldsymbol{D}_{j+1}\vec{v}_{{j+1}}$ and adjust the sign of $\vec{v}_{{j+1}}$ based on
\begin{align}
    \vec{\tilde{v}}_{j+1}=\text{sign} ( \prod_{l=1}^{l=j} r_{l,l+1} ) \vec{v}_{j+1}
       \label{eq:cov_neb_slc}
\end{align}
where $\vec{\tilde{v}}_{j+1}$ represents sign-corrected singular vector from the $(j+1)^\text{th}$ slice.
For the example shown in  Fig.~\ref{fig:method-sign-correction}(a), the correlation between the eigen images from the $1^{\text{st}}$ and $2^{\text{nd}}$ slices, $r_{1,2}$, is negative, which results in sign flipping of $\vec{v}_{2}$ (i.e., $\vec{\tilde{v}}_{2}=-\vec{v}_{2}$) to make it consistent with the reference $\vec{v}_{1}$. For the $3^{\text{rd}}$ slice, in contrast, the sign is not flipped (i.e., $\vec{\tilde{v}}_{3}= \vec{v}_{3}$) since $\prod_{l=1}^{l=2} r_{l,l+1} = r_{1,2}r_{2,3}$ is positive, which indicates that the $3^{\text{rd}}$ slice is already consistent with the reference.
This process is repeated for all values of $j$ as depicted in Fig.~\ref{fig:method-sign-correction}(a). Note, only the pair-wise correlations between two adjacent slices are used for adjusting the sign. The eigen images from non-adjacent slices may not be strongly correlated due to dissimilar image content. As a result, adjusting the sign of $j>2$ slices based on their correlation with $j=1$ slice is not expected to be reliable. Upon the conclusion of this step, the sign of $\vec{\tilde{v}}_{j}$ is consistent across slices but is still arbitrary with respect to the directionality of the respiratory motion and thus cannot distinguish inspiration from expiration. %\textr{RA: Argue that why direct correlation to the reference is not a good idea}.

In the second step, we use the temporal profile of the zeroth-moment center (ZMC) of the image intensity to perform the overall sign correction. The movement of the thoracic and abdominal organs as a function of respiration is predictable in the superior-inferior (SI) direction~\cite{wang1995respiratory}. In particular, ZMC is expected to move superiorly with expiration and inferiorly with inspiration and thus can be used to guide the sign correction. Let $\vec{c}_j \in\Real^{N\times 1}$ represent ZMC across $N$ frames from the $j^\text{th}$ slice. To compute $\vec{c}_j$, we first identify the image axis that has a larger component in the SI direction. The information regarding the precise orientation of the imaging plane is readily available in the Digital Imaging and Communications in Medicine (DICOM) file header. Using this information, we reorient the image such that the vertical axis has the larger component in the SI direction and the top-to-bottom direction of the reoriented image is aligned with superior-to-inferior direction of the subject.
%\textr{(RA: It's not clear how we are estimating LoM along SI. Do we first find LoM curve along one of the image axes then transform it to SI? Also, is there a requirement that one of the image axis has a component along SI? What if both image axes have components along SI? Overall, I think Eqs. 4 and 5 and the text that follows is too concise and hard to follow.)The $\vec{c}_j$ curve for the $j^{\text{th}}$ slice is obtained by tracking the ZMC along the image axis that has a larger component in the superior-inferior (SI) direction. Since the orientation of the image is arbitrary, as shown in Fig.~\ref{fig:method-sign-correction}(b), first we rotate the images so that the vertical axis has the larger component in the SI direction and the superior is on the top.}
 Then, we construct the SI projection $\boldsymbol{P}_j \in\Real^{L\times N}$ of the image series $\boldsymbol{I}_j \in\Real^{R\times L\times N}$ by summing the $R$ pixels along the horizontal direction, where $L$ is the number of pixels in the vertical axis. For the $m^{\text{th}}$ pixel in the $i^{\text{th}}$ frame (column) of $\boldsymbol{P}_j$, its distance from ZMC is defined as: $d(m,i,j) =  \sum_{l=1}^{m}\boldsymbol{P}_{j}(l,i) - \sum_{l=1}^{L}\boldsymbol{P}_{j}(l,i)/2$. 
We find $m^*\geq$ ZMC that minimizes this displacement, i.e.,
 \begin{align}
m^* =\text{argmin}_m d(m,i,j) \quad \text{s.t.} \quad  d(m,i,j)  \geq 0, \quad m \in \{1,2,...,L\}.
\end{align}
 Finally, the ZMC for the $i^{\text{th}}$ frame and $j^{\text{th}}$ slice is obtained by:
 \begin{align}
\vec{c}_{j}(i) &= m^* - \frac{d(m^*,i,j)}{d(m^*,i,j) - d(m^*-1,i,j)}, 
\end{align}
where $d(m^*,i,j) \geq 0$, $d(m^*-1,i,j) < 0$, and $\vec{c}_{j}(i) \in (m^*-1, m^*]$ is the linearly interpolated zero crossing of $d(m,i,j)$.

% Therefore, $\vec{c}_j$ calculated from the image series can be used as a surrogate for the respiratory motion. Then, 
The collective sign of $\vec{\tilde{v}}_{j}$ is adjusted by 
\begin{align}
    \vec{\hat{v}}_{j}=\text{sign}\left[\sum_{j'=1}^{J} \text{sign}(s_{j'})\text{max}(|s_{j'}|-\tau,0)\right]\vec{\tilde{v}}_{j},
    \label{eq_overal_sign}
\end{align}
where $s_{j}$ represents Pearson correlation coefficient between $\vec{\tilde{v}}_{j}$ and $\vec{c}_j$, $\vec{\hat{v}}_{j}$ represents the respiratory signal from the $j^\text{th}$ slice after the overall sign correction, and $\tau = 0.7$ is the correlation coefficient threshold.
In Eq. \ref{eq_overal_sign}, we sum over the weighted sign of correlations, with weights equal to $\text{max}(|s_{j}|-\tau,0)$, which effectively discard slices with low correlation. Then, the overall sign is based on the sign of this sum.
Note, the quality of $\vec{c}_j$ and thus the correlation of $\vec{c}_j$ with $\vec{\tilde{v}}_{j}$ can be poor for certain slices. As a result, this step by itself cannot be reliably used to adjust the sign for individual slices.% is related to the physical motion directly based on the image orientation in image header, so $\vec{\hat{v}}_{j,2}$ is consistent with the respiratory motion and can be used to identify the respiratory phase of each heartbeat.

\section{METHODS}
\subsection{MR acquisition}    
Our method was evaluated in vivo using data from eleven healthy volunteers and two patients. For the recruitment and consent of human subjects, the ethical approval was given by an Internal Review Board (2005H0124) at The Ohio State University. Two ten-slice short-axis stacks of RT cine covering the whole heart were acquired for each volunteer on a 1.5 T scanner (MAGNETOM  Avanto, Siemens Healthcare, Erlangen, Germany), one under free-breathing and the other under end-expiration BH conditions, respectively. The BH acquisition was performed as a reference for cardiac function quantification. 
The acquisition time for each slice was 10 s to cover 1-3 respiratory cycles and 8-14 heartbeats. The slices were acquired sequentially with ECG trigger on. There was a small time gap (less than 1 s) between slices, and the starting respiratory phase for each slice was unknown and arbitrary. The other imaging parameters include: flip angle $69^{\circ}$, TE/TR $1.18/2.83$\,ms, pseudo-random Cartesian sampling with acceleration rate of 9~\cite{VISTA}, slice thickness 8 mm, interslice gap 2 mm, FOV $\sim 300\times400$\,mm, spatial resolution $1.6\times1.6$\,mm$^2$, and temporal resolution $40$--$45$\,ms, resulting in 222--250 frames per slice.
% Each slice was continuously acquired for 10 s to cover 1-3 respiratory cycles and 8-14 heartbeats. The slices were acquired sequentially, with a small (less than one second) time gap between slices. The starting respiratory phase for each slice was arbitrary and unknown. The other imaging parameters were: slice thickness 8 mm, space between adjacent slices 2 mm, FOV $\sim 300\times400$\,mm, TE/TR $1.18/2.83$\,ms, image matrix $192\times256$, flip angle $69^{\circ}$, acceleration rate $9$ using pseudo-random variable density sampling, spatial resolution $\sim 1.6\times1.6$\,mm$^2$, and temporal resolution $40$--$45$\,ms , resulting in 222--250 frames/slice.
The two patients were scanned using the same protocol, with slightly worse temporal ($\sim 51$ ms) and spacial ($\sim2\times2$ mm$^2$) resolutions on a 1.5 T scanner (MAGNETOM  Sola, Siemens Healthcare, Erlangen, Germany) under free-breathing conditions. All the images were reconstructed inline using the Gadgetron~\cite{hansen2013gadgetron} implemention of a parameter-free compressed sensing method called SCoRe~\cite{chen2019sparsity}. %, which is a parameter-free compressed sensing reconstruction method for dynamic MRI.

\subsection{Evaluation}
\begin{figure}[htb]
  \centering
  \centerline{\includegraphics[width=12cm]{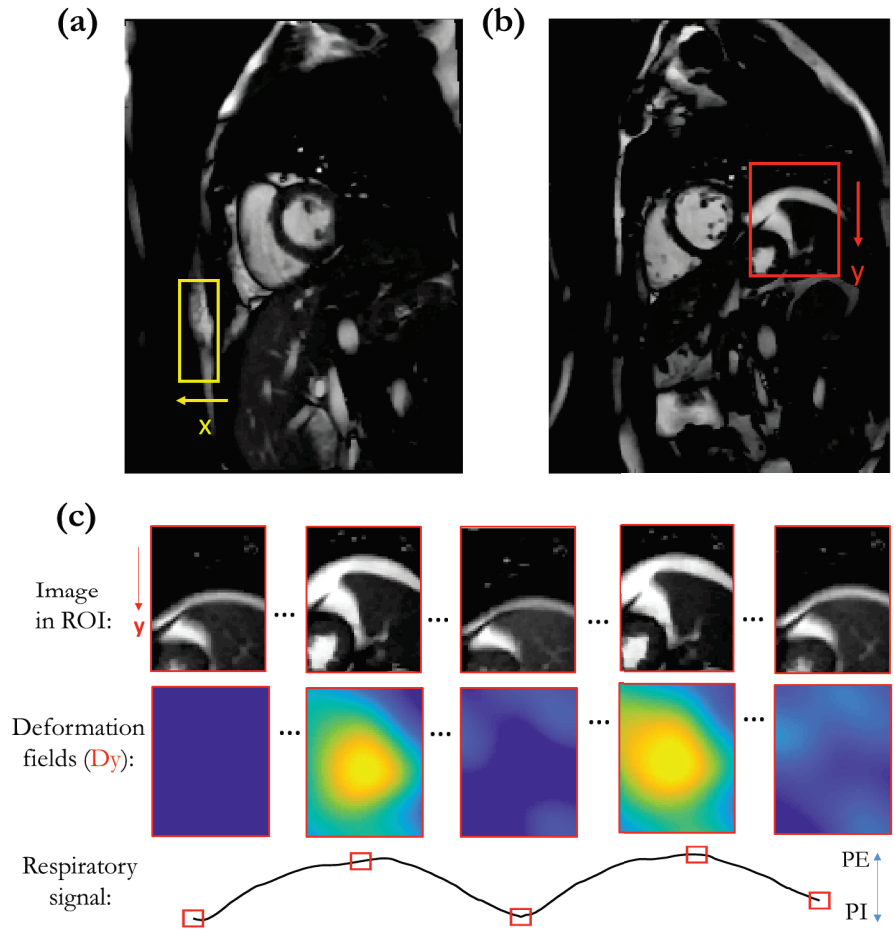}}
%  \vspace{2.0cm}
\caption{ (a, b) Manually placed ROI (red and yellow boxes) for two different volunteers. (c) The pipeline to extract the respiratory motion using non-rigid image registration. Five different frames are shown for the image in (b), along with the y-component of the deformation field, $\text{D}_\text{y}$. The spatial average of $\text{D}_\text{y}$ yields the respiratory signal.}
\label{fig:res}
\end{figure}

As shown in Fig.~\ref{fig:res}, the respiratory motion extracted by non-rigid image registration~\cite{chefd2002flows} within a manually selected region-of-interest (ROI) was used as a reference. For each slice, the ROI was placed in an area with visibly pronounced respiratory motion. To capture the frame-to-frame respiratory motion, all frames for a given slice were registered to the first frame to produce 2D deformation fields with horizontal and vertical components. 
%The motion within a manually selected region-of-interest (ROI) was used as a reference for the respiratory motion. As shown in Fig.~\ref{fig:res}, for each slice, the ROI was placed manually in an area with visually pronounced respiratory motion. The frame-to-frame motion was captured using non-rigid image registration~\cite{chefd2002flows}. To this end, all frames for a given slice were registered to the first frame to produce 2D deformation fields with horizontal and vertical components.
%The deformation field characterized the motion of each pixel across $N$ frames.  The respiratory signal was extracted by averaging the horizontal or vertical component of the deformation field within the ROI. 
One of the components was averaged over the ROI to yield the reference signal across $N$ frames. The appropriate component of the deformation field was selected based on the visual assessment of the motion within the ROI. 
For example, in Fig.~\ref{fig:res}(a), the ROI was selected on the chest wall, where the respiratory motion was predominantly along the anterior-posterior (AP) direction in the patient coordinate system. For this ROI, we chose the horizontal component of the deformation field since it aligns more closely with the AP direction. However, in Fig.~\ref{fig:res}(b), the ROI was selected on the liver dome, which moves along the SI direction. For this ROI, the vertical component of the deformation field was selected.
To evaluate the agreement between the extracted respiratory signal, $\vec{\hat{v}}_{j}$, and the reference, Pearson correlation coefficients were calculated for each slice. The directionality determination was considered to be correct when the correlation was positive.
%Agreement between the respiratory signal from the proposed method, $\vec{\hat{v}}_{j}$, and the reference was evaluated using correlation coefficients. When the correlation was positive, the directionality determination was considered to be correct.

As a comparison, we also implemented URE~\cite{MRIPCA}, where the lung-diaphragm interface is selected automatically using the eigen image of the middle slice of the short-axis stack; then, the directionality of the respiratory signal is determined by tracking the diaphragmatic motion or estimating the lung volume. 
Using multi-slice RT cine data from eleven volunteers and two patients, we compared our method to URE to extract the respiratory signal. For each method, all instances where the final respiratory signal had negative correlation with the reference were considered unsuccessful. Both methods were implemented Matlab (Mathworks, Natick, Massachusetts). The code of our method is provided on Github: \url{https://github.com/MRIOSU/AERC}.%The metric we use to evaluate those two methods is the number of the incorrect slices, whose extracted respiratory signals are negatively correlated with the reference. The fewer number of incorrect slices, the better the method performs.}
%\textr{(How is the comparison made? What is the criteria for success? Also, we also need to mention what we did with the data from two patients.)}%Our method is compared with ~\cite{MRIPCA} for all the subjects.
% the deformation field along the visually determined orientation of the respiratory motion. Similar to the COM curve, the sign of the reference with respect to the physical motion is determined using the information in the image header.

\subsection{Cardiac function quantification}  
Using the extracted respiratory signal as a guide, we selected one peak-expiration (PE) and one peak-inspiration (PI) heartbeat from each slice, and showed the respiratory motion-induced variations in cardiac output for healthy subjects. The quantification from the end-expiratory BH scan was chosen as the reference. For RT processing, arrhythmic heartbeats, defined as having a duration greater than fifteen percent differnt from the mean RR interval, were discarded from consideration.
Short-axis contours were initially generated automatically using the suiteHEART software (NeoSoft LLC, Pewaukee, WI) and then were visually assessed frame-by-frame and modified manually when necessary.  
In this study, trabeculations and papillary muscles were considered part of the right and left ventricular blood pools. After the summation of all slices, suiteHEART calculated the end-diastolic volume (EDV) and end-systolic volume (ESV)
automatically for each ventricle. The analysis was firstly performed in volunteers for the PE heartbeats, and then repeated for the PI heartbeats using the same protocol. %Stroke volume (SV) was calculated as $\text{EDV}-\text{ESV}$.
%, and ejection fraction (EF) was defined as $\text{SV}/\text{EDV}$.

\section{RESULTS}
\subsection{Performance in healthy volunteers}
\begin{figure}[htb]
  \centering
  \centerline{\includegraphics[width=16cm]{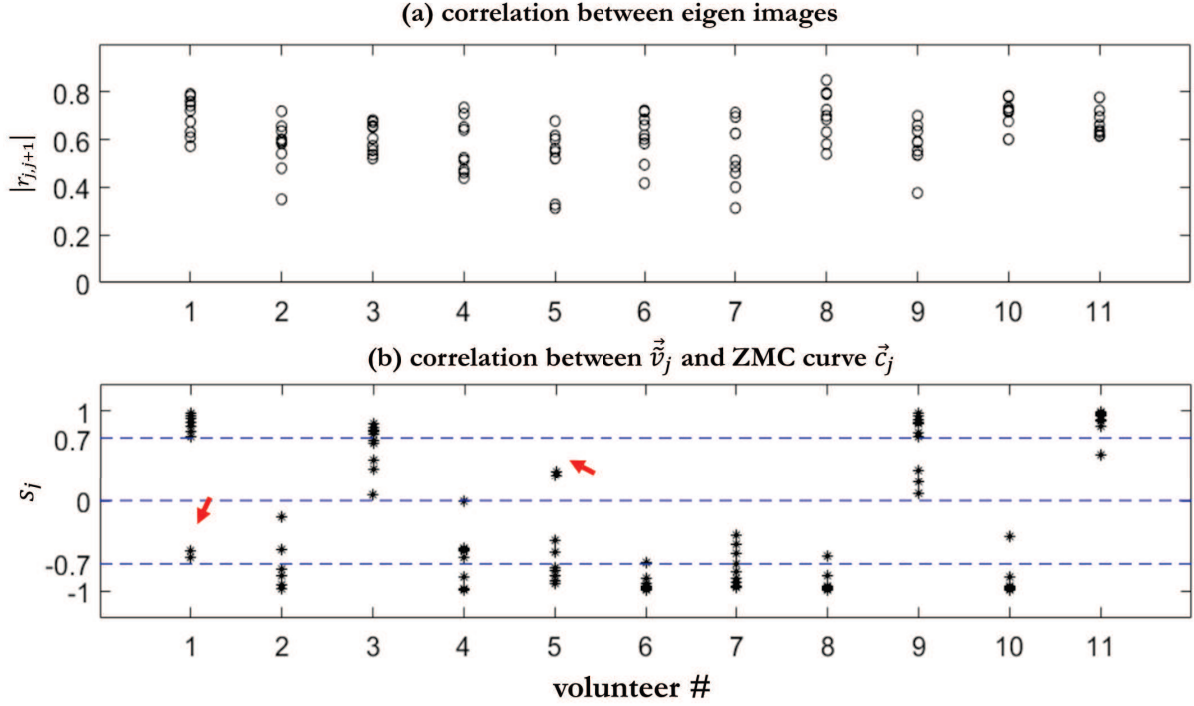}}
\caption{ (a) The absolute values of Pearson correlation coefficients $\left |r_{j,j+1}\right |$ between eigen-images, i.e.,  $\boldsymbol{D}_j\vec{v}_{j}$ and $\boldsymbol{D}_{j+1}\vec{v}_{{j+1}}$, which are used in the first step of our directionality determination procedure. (b) Pearson correlation coefficients $s_{j}$ between $\vec{\tilde{v}}_{j}$ and ZMC curve $\vec{c}_j$ 
 used for overall sign correction in the second step. The red arrows highlight instances where the signs of individual slices are incorrect.}%The sign of  $\vec{\tilde{v}}_{j}$ has been adjusted with respect to the reference slice in the previous step.}
\label{fig:cov_values}
\end{figure}

% \textr{First describe Figure 5, then Table 1, then Figure 4, and then Figure 6.}
Fig.~\ref{fig:cov_values}(a) shows the absolute values of Pearson correlation coefficients $\left |r_{j,j+1}\right |$ between neighboring eigen images, $\boldsymbol{D}_j\vec{v}_{j}$ and $\boldsymbol{D}_{j+1}\vec{v}_{{j+1}}$, for all the volunteers. The mean and standard deviation of $\left |r_{j,j+1}\right |$ is 0.65 $\pm$ 0.16. %, which is not large as expected. However, it is sufficiently  reliable to adjust the sign of all slices with respect to the reference. 
Fig.~\ref{fig:cov_values} (b) shows the Pearson correlation coefficients, $s_{j}$, between $\vec{\tilde{v}}_{j}$ and ZMC curve, $\vec{c}_j$. As highlighted by the red arrows, the sign of $s_{j}$ is incorrect for some individual slices due to the poor quality of the ZMC curve. Therefore, the ZMC curve itself cannot be used to determine the sign for individual slices. However, the overall sign correction using Eq. \ref{eq_overal_sign} is robust because it is based on consensus from all slices that exhibit strong correlation between $\vec{c}_j$ and $\vec{\tilde{v}}_{j}$.

\begin{figure}[htb]
  \centering
  \centerline{\includegraphics[width=16cm]{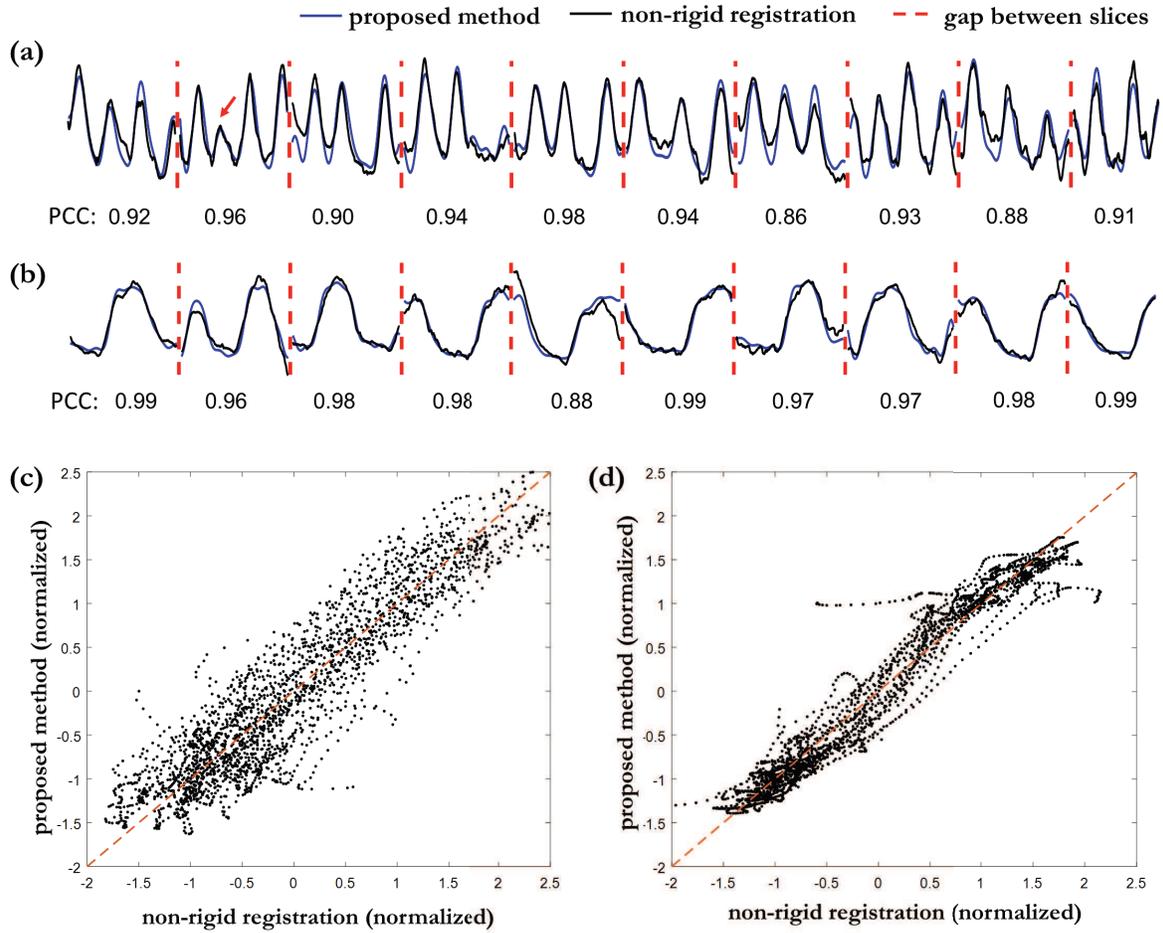}}
\caption{ (a, b) The respiratory signal from the proposed method and the reference superimposed. (c, d) Scatter plots between the extracted and reference respiratory signals in (a) and (b). The results from two representative volunteers are shown, one with short respiratory cycles (a, c) and the other with long cycles (b, d). The red arrow highlighting the irregular respiratory motion. PCC is the Pearson correlation coefficient between signals from the proposed method and the reference.}
%\textr{RA: for (c, d) remove capitalization to make it consistent.} 
\label{fig:res_curve}
\end{figure}

Pearson correlation coefficients between the extracted respiratory motion and the reference for healthy subjects are summarized in Table \ref{tab:table_R}. We report the mean, standard derivation (SD), median, and range across 10 slices for each subject. In all cases, the extracted respiratory signal exhibit high, positive correlation with the reference. Fig.~\ref{fig:res_curve} shows results from two representative volunteers, one with short respiratory cycles and the other with long cycles. The proposed method works well across different respiratory patterns. As highlighted by the red arrow, this method is also tolerant of irregular respiratory motion. For some slices in Fig.~\ref{fig:res_curve}, the discrepancy between the proposed method and reference is more pronounced. This is not surprising because the two methods measure different quantities: the reference (non-rigid image registration) extracts the respiratory motion from a  user-defined ROI, while the proposed PCA-based method tracks the motion from the entire image.

%\textr{Table \ref{tab:table_R} summarizes the Pearson correlation coefficients between the extracted respiratory motion and the reference for healthy subjects. We report the mean, standard derivation (SD), median, and range across 10 slices for each subject. The results show that the extracted respiratory signal has a high, positive correlation with the reference in all cases. This method is also robust across different respiratory patterns.  Fig.~\ref{fig:res_curve} shows representative results from two volunteers, one with short respiratory cycles and the other with long cycles. As highlighted by the red arrow, this method is also tolerant of irregular motion. For some slices in Fig.~\ref{fig:res_curve}, the difference between the reference and the proposed method is more pronounced. This is expected because the two methods measure different quantities: the proposed PCA-based method extracts the respiratory motion from the entire image, while the non-rigid image registration only tracks the motion in a small user-defined ROI. }
%\subsection{Reliability}

\begin{figure}[htb]
  \centering
  \centerline{\includegraphics[width=15cm]{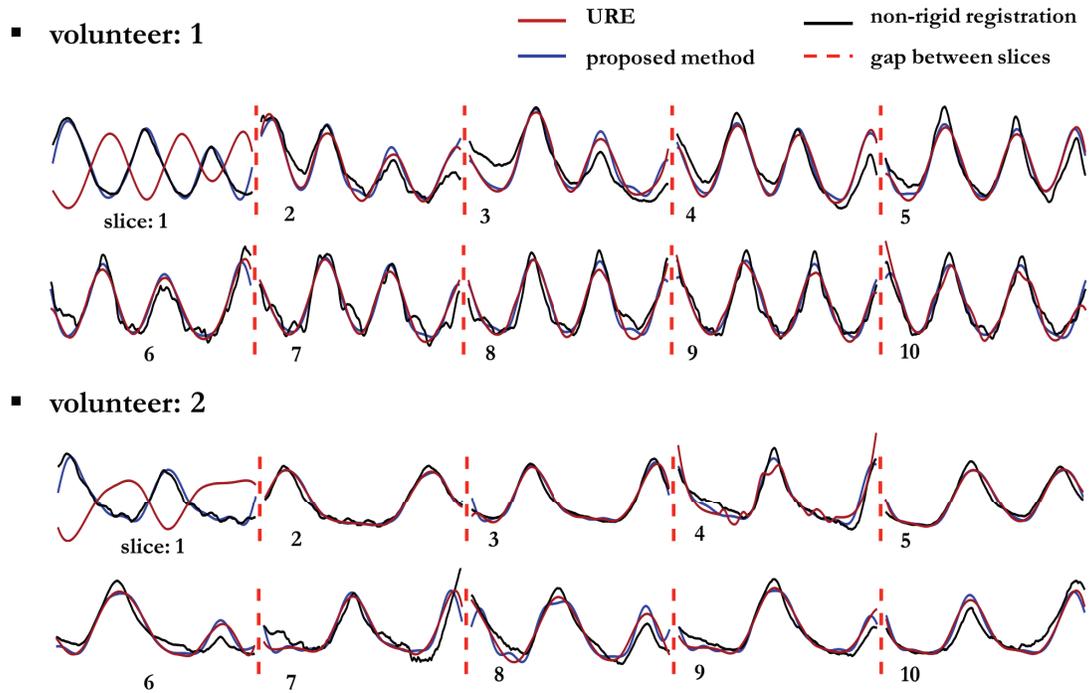}}
\caption{ A direct comparison of the respiratory signals extracted using the proposed method and the method by Novillo et al. (URE). The results are shown for two volunteers. URE determined the directionality of the respiratory signal correctly for all the other volunteers, but failed in the two slices shown here.}
\label{fig:compare}
\end{figure}

% \subsection{Comparison with the previous method}
Fig.~\ref{fig:compare} shows the respiratory signals extracted using our method, URE, and the non-rigid registration reference for two volunteers. URE determined the directionality of the respiratory signal correctly for all the other volunteers but failed in two slices shown here. One of the failures was caused by the wrong selection of lung-diaphragm interface, and the other by the incorrect directionality of the signal estimated from the lung volume. In contrast, our method corrected the directionality of the respiratory signal successfully for all the cases. We also observed that there are more cardiac motion-induced oscillations in the non-rigid registration reference in Fig.~\ref{fig:compare}; it is expected because no filter is applied during the extraction of the reference signal. The averaged processing time (Intel Core i5, 2.7 GHz, 16 GB) for one volunteer using our method and URE was 10.0 s and 13.4 s, respectively.

%Using the extracted respiratory signal as a guide, we extracted one peak-inspiration (PI) and one peak-expiration (PE) heartbeat from each slice.
\subsection{Heartbeat selection and cardiac function quantification}
\begin{figure}[htb]
  \centering
  \vspace{0.1cm}  
  \centerline{\includegraphics[width=16cm]{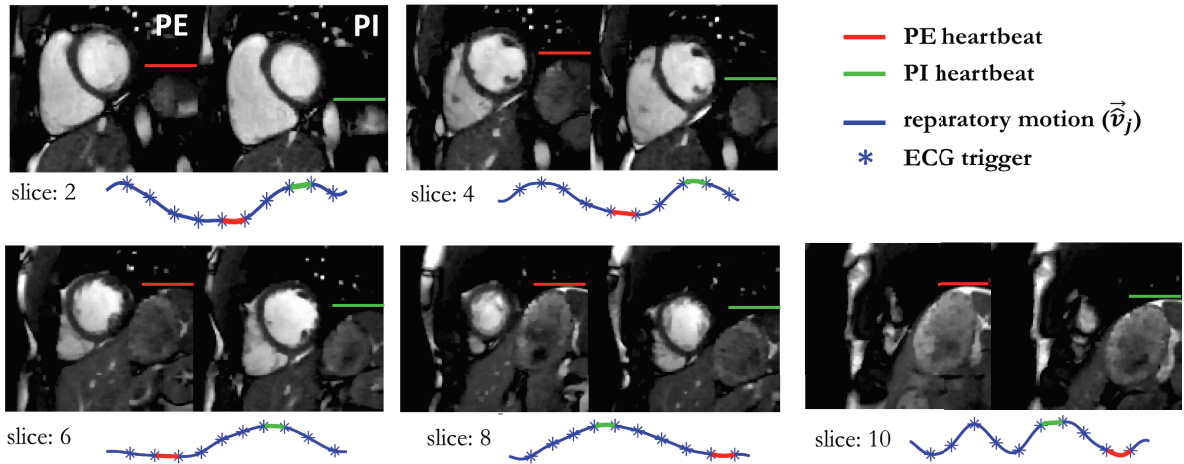}}  
 % \vspace{0.3cm}
   % \vspace{0.2cm}  
 % \centerline{\includegraphics[width=8.5cm]{Result2_EI.PNG}}  
\caption{Peak expiration (PE) and peak inspiration (PI) heartbeats identified using the proposed method. Images from five different slices are shown. The position of the liver dome is tracked with red (PE) and green (PI) lines. As expected, the position of the liver dome is lower in PI than in PE. The through plane motion between PE and PI is also obvious from the size of the RV. Only an end-diastolic frame is shown for one volunteer.}
%\textr{RA: Remove the capitalization from "Respiratory", "Trigger" and "Slice". Also, insert a space after "slice:". For example, change "Slice:2" to "slice: 2"}
\label{fig:res_image}
\end{figure}

\begin{figure}[htb]
  \centering
  \vspace{0.1cm}  
  \centerline{\includegraphics[width=18cm]{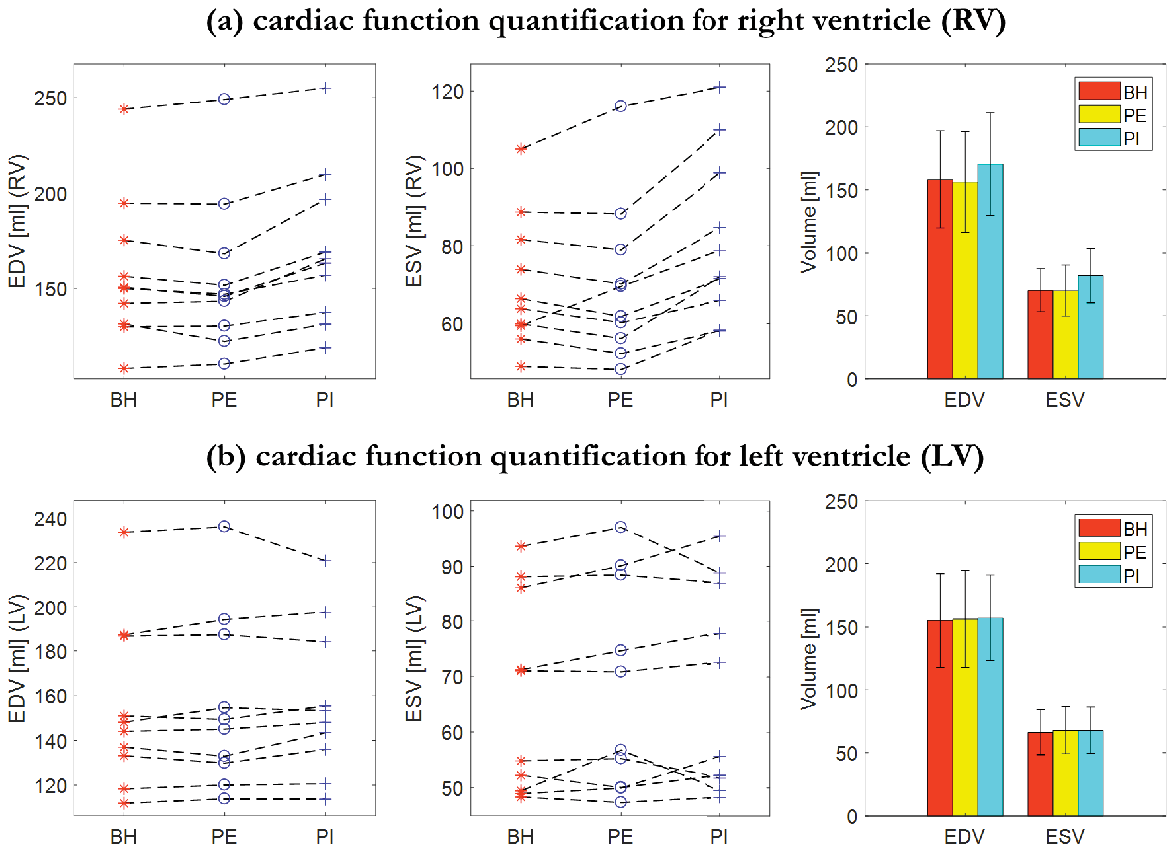}}  
\caption{Cardiac function quantification for the right (a) and left (b) ventricle. EDV and ESV are reported here. Each line in the figure represents the result from a single volunteer. The bar chart shows the mean and standard deviation of EDV and ESV across all the volunteers. }
\label{fig:cardiac_quant}
\end{figure}

Fig.~\ref{fig:res_image} shows the PE and PI heartbeats selected using the extracted respiratory signal as a guide for one volunteer. In the end-diastolic frames, all slices within PE have elevated liver dome compared to the corresponding PI slices. Significant in-plane and through-plane motion of the heart can be noticed between PE and PI heartbeats, highlighting the perils of performing quantification from slices with disparate respiratory phases. Fig.~\ref{fig:cardiac_quant}  summarizes the results of cardiac function quantification for both ventricles. EDV, ESV, their mean and standard deviation across ten volunteers are reported here. For one volunteer, there was visible misalignment between the BH and free-breathing scans, potentially due to the patient motion; this subject was excluded from cardiac function quantification. For each of the remaining ten volunteers, quantification results from BH, PE, and PI are reported, with BH serving as reference. PE and PI represent quantification performed from peak-expiration and peak-inspiration heartbeats. As shown in  Fig.~\ref{fig:cardiac_quant}(a), for the right ventricle, there is a significant increase (p < 0.01) from PE to PI both in EDV and ESV, and the quantification of PE is in agreement with reference BH. 
These findings are consistent with the literature \cite{claessen2014interaction}.
However, for the left ventricle, the differences between BH, PE, and PI are not significant, which is likely due to the sample size being inadequate to distinguish smaller variations in left ventricle (LV) function \cite{claessen2014interaction,liu2019extending}.
%\textr{(These findings are consistent with the literature(?)}.

\subsection{Performance in patients}
Fig.~\ref{fig:compare} demonstrates the respiratory signals extracted using our method, URE and the non-rigid registration reference for the two patients. All the methods performed reasonably well in the second patient, who had a periodic respiratory pattern. However, for the first patient with irregular breathing motion, the respiratory signal extracted using the proposed method is in better agreement with the non-rigid registration reference.

%To illustrate the ability of identifying the end-expiration (PE) and end-inspiration (PI) heartbeats using the proposed method, we averaged the respiratory displacement in each heartbeat, and choose the one with minim/maximal displacement as PE/PI heartbeat. Figure. \ref{fig:res_curve} shows the selected PE/PI heartbeats, as well as the corresponding end diastolic frames. As highlighted by the blue arrows and red lines, both in-plane and through-plane motion of the heart can be noticed, which will lead to different cardiac output.

\section{DISCUSSION}
To accurately and reproducibly measure LV and right ventricle (RV) cardiac function, it is essential to determine the respiratory phase of the images. The feasibility of dimensionality reduction-based methods to extract the respiratory signal has been previously demonstrated. However,  none of these methods preserve the directionality of respiratory signal, and therefore cannot reliably distinguish between inspiratory and expiratory motion, which can lead to inaccurate cardiac function quantification. We propose and evaluate a new method to correctly assign the directionality (sign) of the respiratory signal where no ROI and tuning parameters are induced. The method utilizes the similarity of neighboring eigen images to adjust the sign of all slices with respect to the first slice. Then, the zeroth-moment center, ZMC, is used to adjust the global sign. In this work, linear PCA is used to extract the respiratory signal; however, the proposed directionality determination procedure can be incorporated with other dimensionality reduction-based methods for respiratory signal extraction. %The feasibility of the proposed method sign correction scheme is demonstrated in healthy 11 volunteers and two patients.

In the volunteer study, we evaluated our method in 11 subjects. The respiratory signal extracted using non-rigid image registration from a manually selected ROI is chosen as the reference. To evaluate the reliability of our method, we have shown the absolute values of Pearson correlation coefficients $\left |r_{j,j+1}\right |$ between neighboring eigen images and the Pearson correlation coefficients, $s_{j}$, between $\vec{\tilde{v}}_{j}$ and the ZMC curve $\vec{c}_j$ in Fig. \ref{fig:cov_values}. The content of neighboring slices is expected to be similar but not identical; hence, the values of $\left |r_{j,j+1}\right |$ are significantly less than one. However, our preliminary findings show that these values are sufficiently reliable to adjust the sign of all slices with respect to the reference slice. After the first step of our directionality (sign) determination procedure, the sign of $\vec{\tilde{v}}_{j}$ is consistent for all values of $j$. If the ZMC curve is accurate, $s_{j}$ should have the same sign, either positive or negative, for all $j$. However, as highlighted by the red arrows in Fig.\ref{fig:cov_values}(b), the sign of $s_{j}$ is incorrect for certain slices due to the poor quality of $\vec{c}_j$, which means the ZMC curve by itself cannot be used to determine the sign for individual slices. %Hence, methods based on only Step 2 are not expected to perform reliably. 
As shown in Fig.~\ref{fig:res_curve}, the proposed method is robust across different respiratory patterns, even for irregular respiratory motion. Since the reference only tracks the motion in the selected ROI, while the proposed method extracts the respiratory motion from the entire image, the difference between them is pronounced for some slices. Overall, our method determined the directionality of the respiratory signal successfully in all cases. 

Although linear PCA was used to extract the respiratory signal from RT cardiac cine MRI both in this work and in URE~\cite{MRIPCA}, there are minor differences between the two approaches: (i) In our method, the images are pre-filtered to suppress the cardiac motion and (ii) the temporal mean is subtracted before applying the linear PCA. Consequently, our largest singular vector consistently captures the respiratory motion. In contrast, the first singular vector in URE is the DC component and the second singular vector is the respiratory signal, which is then filtered using a Savitzky–Golay filter to remove cardiac signal. With the cardiac motion filtered and the temporal mean subtracted before PCA, our method is more robust in localizing the respiratory motion to the first singular vector. Without these two additional processing steps, it is feasible that the temporal mean, respiratory motion, and cardiac motion are not localized in the first, second and third singular vectors, respectively, leading to contamination of the extracted respiratory signal. Our directionality determination procedure, which is the major contribution of this work, is distinct from URE. URE determined the directionality of the respiratory signal by tracking the diaphragmatic motion or estimating the lung volume, where the lung-diaphragm interface is automatically identified. However, in this work, we propose a two-step procedure to perform the directionality determination, where no ROI and tuning parameters are induced.
%In Figure.6, Novillo's algorithm determined the sign of the respiratory signal correctly for all the other volunteers but failed in the two basal slices shown here. In contrast, our method determined the sign of the respiratory signal successfully for all cases. 
%Besides the sign correction, we extract the respiratory signal using a different implementation: (i) The images were filtered to kill the cardiac motion at the beginning; (ii) The temporal mean is subtracted before applying the linear PCA. However, in Ref.~\cite{MRIPCA}, there is no filtering and temporal mean subtraction. 

\begin{figure}[htb]
  \centering
  \centerline{\includegraphics[width=15cm]{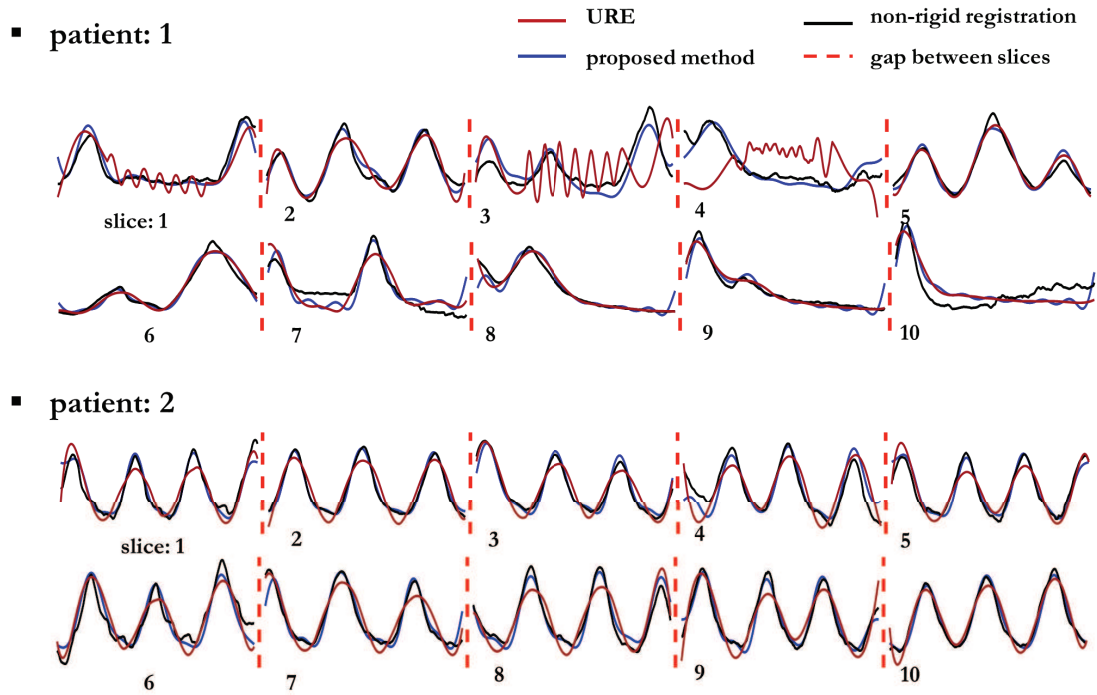}}
\caption{A direct comparison of the respiratory signals extracted using the proposed method and the method by Novillo et al. (URE). The results are shown for two patients.}
\label{fig:compare_patient}
\end{figure}

To demonstrate the clinical utility of our method, we selected PE and PI heartbeats from each slice using the extracted respiratory signal as a guide. As shown in Fig.~\ref{fig:res_image}, both significant in-plane and through-plane motion can be noticed between the PE and PI heartbeats. Several studies have investigated the influence of breathing in left and right ventricular volumes. During normal breathing, inspiration results in a significant increase in RV volumes and a small decrease in LV volumes. In this work, we observed a significant increase ($p < 0.001$) from PE to PI both for RV EDV and RV ESV, which is consistent with the the physiology~\cite{claessen2014interaction} and highlight the potential adverse affect of mixing slices from different respiratory phases. 
However, for the LV volume, we did not observe a significant difference between the PE and PI heartbeats. This could be because the small sample size is inadequate to distinguish variations in LV function, which is less susceptible to respiratory-induced motion~\cite{liu2019extending}.
We also compared the two methods using data from two patients, one with a periodic respiratory pattern and the other with an irregular respiratory pattern. As shown in Fig. \ref{fig:compare_patient}, both methods performed well when the respiratory pattern was periodic; however, for the patient with irregular breathing motion, our method exhibited better agreement with the reference.

We recognize that the proposed method has several limitations. One of them is the expectation that the respiratory signal is confined to the first singular vector. 
Besides the eleven subjects included in this study, we collected data from one more volunteer. Due to the significant bulk motion during the scan, this volunteer was excluded from the study. Nonetheless, we observed that the first singular vector, for this specific volunteer, captured the bulk motion instead of the respiratory motion.
As it has been reported previously~\cite{MRIPCA}, the presence of another motion source, can threaten the localization of the respiratory motion to one singular vector. 
Another limitation of this study is the lack of validation for the cases with shorter scan time ($< 10$\,s). In this work, all the subjects were scanned for 10\,s per slice to cover 1-3 respiratory cycles. Since both the low pass filter and PCA benefit from the long scan, we expect that both the performance of our method and URE to degrade for shorter scans. In order to explore the influence of shorter scan time, we separated each 10\,s cine series into two 5\,s series.
%We effectively reduced the scan time by keeping the images of the first 5 and 2.5 s for each slice to explore its influence on the performance of the proposed method. 
We found that the directionality in 1 and 3 cases were incorrectly determined for 5 s series using the proposed method and URE, respectively. Since the proposed method utilizes the correlation of neighboring eigen images (Eq.~\ref{eq:cov_neb_slc}), the incorrect directionality assignment for the $l^{\text{th}}$ slice may transmit to the following slices ($j > l$). We did not encounter this problem in the 10\,s and 5\,s series, but this is a potential risk for shorter scan times. To accommodate this potential limitation, non-rigid registration could be applied on neighboring eigen images to increase their correlation from a shorter scan. %In contrast, the 10 s scan allows correct directionality assignment for all cases. % and respiratory signal for more slices was assigned incorrectly in the truncated images, compared to the original 10 s/slice images: 12 slices for 2.5 s, 1 slice for 5 s whereas no failures for 10 s. %As expected, when the scan time was reduced to 5 and 2.5 s/slice effectively, the proposed method failed to determine the sign of the respiratory signal correctly for 1 (1) and 12 (4) slices (volunteers) across 110 (11)  slices (volunteers) respectively.
Finally, this study does not include a large number of subjects with inconsistent breathing. %In the future, we will evaluate the proposed method in a larger population that includes subjects with inconsistent breathing patterns.
%\textb{and only investigate the extraction of the respiratory signal at rest. , and its ability to extract the respiratory signal during or right after exercise.}
% %We will also apply this method to the subjects scanned under exercise, which is much more challenging compared to the scan under rest.

%Along with PCA, we can accurately identify heartbeats, one from each slice, belonging to the same respiratory phase, which can improve cardiac function  quantification  for  RT  CMR  while  eliminating  the need to manually select heartbeats from each slice.

 %The proposed method worked very well for all the scans, even when the gap between neighboring slices is doubled. However, there is a risk when the gad between neighboring slices is too big, which will lead to very weak correlation between neighboring slices. This can be compensated by shifting one of the neighboring eigen images a little before calculating their correlation coefficients, and finding the amount of shift which leads to the strongest correlation. Another potential issue for this study is related to PCA, where the respiratory motion is derived from the second singular vector. However, it may fail if other kind of motion has more contribution, such as patient movement. This has been noticed in the previous study~\cite{novillo2019unsupervised} and more investigation needs to be performed to address this issue.

%\vfill
%\pagebreak
\section{CONCLUSION}
We have proposed a fully automatic method to extract the respiratory signal and to correctly identify the directionality of the respiratory phase in each slice. Linear PCA is applied to the preprocessed temporal image series to extract the respiratory motion. Then a two-step procedure is used to determine the directionality (sign) of the respiratory signal: (i) utilize the similarity of neighboring eigen images to adjust the sign of all slices with respect to the first slice; (ii) use the ZMC curve to perform the global sign correction. The proposed method was used to extract and determine the directionality of the respiratory signal successfully for 11 volunteers and 2 patients. 
Using the extracted respiratory signal as a guide, we selected one PE and PI heartbeats from each slice and demonstrated the respiratory motion induced variations in cardiac output in volunteers. The proposed method requires no user-defined ROI or tuning parameters and was shown to be more reliable than another recently described method.

\section{ACKNOWLEDGMENTS}
This work was funded in part by NIH project R01HL135489.

\bibliography{references.bib}

\clearpage
%%%%%%%%%%%%%%%%%%Table%%%%%%%%%%%%%%%%%%%%%%%%%%%%
\begin{table}[!htbp]
		\begin{center}
\begin{tabular}{cccc}
\toprule
\multicolumn{4}{c}{{\bf Pearson correlation coefficient}} \\
\midrule
    {Volunteer}     & {Mean$\pm$SD}   & {Median} & {Range} \\ \hline
   {\bf 1} &       0.95$\pm$0.02      & 0.96     & 0.90\textup{--}0.97 \\

   {\bf 2} &       0.91$\pm$0.03      & 0.91     &0.87\textup{--}0.95 \\

   {\bf 3} &       0.94$\pm$0.04      & 0.95     &0.87\textup{--}0.97 \\

   {\bf 4} &       0.93$\pm$0.10      & 0.96     &0.66\textup{--}0.99 \\

   {\bf 5} &       0.92$\pm$0.05      & 0.93     &0.81\textup{--}0.98 \\

   {\bf 6} &       0.93$\pm$0.04      & 0.95     &0.86\textup{--}0.97 \\

   {\bf 7} &       0.92$\pm$0.04      & 0.93     &0.86\textup{--}0.98 \\

   {\bf 8} &       0.93$\pm$0.05      & 0.95     &0.82\textup{--}0.99 \\

   {\bf 9} &       0.92$\pm$0.08      & 0.94     &0.73\textup{--}0.98 \\

  {\bf 10} &       0.96$\pm$0.03      & 0.98     &0.90\textup{--}0.99 \\

  {\bf 11} &       0.97$\pm$0.03      & 0.98     &0.88-0.99 \\
\bottomrule
\end{tabular}  
		\end{center}
	\caption{Pearson correlation coefficients between the extracted respiratory motion and the reference. Ten slices were scanned for each volunteer and the correlation coefficient was calculated slice-wise.}
	\label{tab:table_R}
\end{table}
%%%%%%%%%%%%%%%%%%Table%%%%%%%%%%%%%%%%%%%%%%%%%%%%
\clearpage

%%%%%%%%%%%%%%%%%%%Figure Method%%%%%%%%%%%%%%%%%%
%%%%%%%%%%%%%%%%%%%Figure Method%%%%%%%%%%%%%%%%%%
\clearpage

%%%%%%%%%%%%%%%%%%%Figure Method%%%%%%%%%%%%%%%%%%
%%%%%%%%%%%%%%%%%%%Figure Method%%%%%%%%%%%%%%%%%%
\clearpage

%%%%%%%%%%%%%%%%%%%Figure_ROI%%%%%%%%%%%
%%%%%%%%%%%%%%%%%%%Figure_ROI%%%%%%%%%%%
\clearpage

%%%%%%%%%%%%%%%%%%%%%%figure cov_values%%%%%%%%%%%%%%%%%%%
%%%%%%%%%%%%%%%%%%%%%%figure cov_values%%%%%%%%%%%%%%%%%%%  
\clearpage

%%%%%%%%%%%%%%%%%%%%%%figure Result curve%%%%%%%%%%%%%%%%%%%
%%%%%%%%%%%%%%%%%%%%%%figure Result curve%%%%%%%%%%%%%%%%%%%    
\clearpage

%%%%%%%%%%%%%%%%%%%%%%figure compare%%%%%%%%%%%%%%%%%%
%%%%%%%%%%%%%%%%%%%%%%figure compare%%%%%%%%%%%%%%%%%%%   
\clearpage

%%%%%%%%%%%%%%%%%%%%%%figure Result Image%%%%%%%%%%%%%%%%%%%
%%%%%%%%%%%%%%%%%%%%%%figure Result Image%%%%%%%%%%%%%%%%%%%
\clearpage

%%%%%%%%%%%%%%%%%%%%%%figure cardic_quant%%%%%%%%%%%%%%%%%%%
%%%%%%%%%%%%%%%%%%%%%%figure cardic_quant%%%%%%%%%%%%%%%%%%%
\clearpage

%%%%%%%%%%%%%%%%%%%%%%figure Result_patient%%%%%%%%%%%%%%%%%%
%%%%%%%%%%%%%%%%%%%%%%figure Result_patient%%%%%%%%%%%%%%%%%%%  

\end{document}